\documentclass[12pt]{article}
\usepackage{amsmath,amssymb,bm,epsfig}

\renewcommand{\thefootnote}{\fnsymbol{footnote}}

\begin{document}

\title{
\begin{flushright}
\begin{minipage}{0.2\linewidth}
\normalsize
WU-HEP-14-10 \\
EPHOU-14-019 \\*[50pt]
\end{minipage}
\end{flushright}
{\Large \bf 
Natural inflation with and without modulations\\ 
in type IIB string theory 
\\*[20pt]}}

\author{Hiroyuki~Abe$^{1,}$\footnote{
E-mail address: abe@waseda.jp},\ \ 
Tatsuo~Kobayashi$^{2}$\footnote{
E-mail address:  kobayashi@particle.sci.hokudai.ac.jp}, \ and \ 
Hajime~Otsuka$^{1,}$\footnote{
E-mail address: hajime.13.gologo@akane.waseda.jp
}\\*[20pt]
$^1${\it \normalsize 
Department of Physics, Waseda University, 
Tokyo 169-8555, Japan} \\
$^2${\it \normalsize 
Department of Physics, Hokkaido University, Sapporo 060-0810, Japan} \\*[50pt]}

\date{
\centerline{\small \bf Abstract}
\begin{minipage}{0.9\linewidth}
\medskip 
\medskip 
\small
We propose a mechanism for the natural inflation 
with and without modulation in 
the framework of type IIB string theory 
on toroidal orientifold or orbifold. 
We explicitly construct the stabilization 
potential of complex structure, 
dilaton and K\"ahler moduli, where 
one of the imaginary component of complex 
structure moduli becomes light which is 
identified as the inflaton. 
The inflaton potential is generated by the 
gaugino-condensation term which receives the 
one-loop threshold corrections determined 
by the field value of complex structure moduli 
and the axion decay constant of inflaton is enhanced by the 
inverse of one-loop factor. 
We also find the threshold corrections 
can also induce the modulations to the original scalar potential 
for the natural inflation. Depending on these modulations, 
we can predict several sizes of tensor-to-scalar 
ratio as well as the other cosmological 
observables reported by WMAP, Planck and/or 
BICEP2 collaborations. 
\end{minipage}
}

\begin{titlepage}
\maketitle
\thispagestyle{empty}
\clearpage
\tableofcontents
\thispagestyle{empty}
\end{titlepage}

\renewcommand{\thefootnote}{\arabic{footnote}}
\setcounter{footnote}{0}
\vspace{35pt}

\section{Introduction}
Cosmic inflation is the most successful scenario which 
not only explains the current cosmological observations 
but also solves the fine-tuning problems such as the 
horizon and flatness problems at the same 
time. 

The inflation scenarios are mostly classified according 
to the size of the tensor-to-scalar ratio which 
measures the tensor perturbations of the metric in 
our universe. One is the small-field inflation 
scenario which gives the tiny tensor-to-scalar ratio 
due to the flat potential of the scalar field, called 
inflaton. The other scenario we consider 
is the large-field inflation model which gives a sizable 
and measurable tensor-to-scalar ratio. 
Recent data reported by BICEP2 
collaboration~\cite{Ade:2014xna} can be explained 
by dust emission~\cite{Adam:2014bub,Ade:2015tva} 
reported by the joint 
analysis of BICEP2, Keck Array and Planck 
collaborations. 
In any case, it is interesting to propose the 
large-field inflation models which would be tested 
by future cosmological observations. 

When we consider the large-field inflation models, 
we always encounter the problems how to treat the 
trans-Planckian field values. For example, in the case of 
natural inflation~\cite{Freese:1990rb} known as one of 
the large-field models, 
we need the corresponding trans-Planckian axion decay 
constant of the inflaton which is required by recent 
Planck data~\cite{Planck:2013jfk,Ade:2015lrj}. 
(See Ref.~\cite{Freese:2014nla} and references therein.)

Especially, in the higher-dimensional theory, there are a lot of 
axions associated with the internal cycles of the internal 
manifold and then it would be natural to identify such axions as 
the inflaton. 
However, it is in general to be problematic that 
the scale of axion decay constant is severely 
constrained by the size of internal manifold and the cut-off 
scale of higher-dimensional theory. 
To overcome such a problem, there are several approaches to 
realize trans-Planckian axion decay constant by employing 
Kim-Niles-Peloso alignment mechanism~\cite{Kim:2004rp} 
in the case of multiple axions with sub-Planckian axion 
decay constant, for more details see 
Refs.~\cite{Kallosh:2007ig,Czerny:2014xja,Choi:2014rja,
Tye:2014tja,Kappl:2014lra,Ben-Dayan:2014zsa,
Long:2014dta,Li:2014lpa,Kenton:2014gma,Ali:2014mra}. 
In the case of single axion, the trans-Planckian axion decay 
constant can be realized based on the five-dimensional 
theory~\cite{ArkaniHamed:2003wu,Abe:2014vca} with a 
small five-dimensional gauge coupling and the weakly-coupled 
heterotic string theory with certain 
loop-corrections to the gauge coupling~\cite{Abe:2014pwa}. 
 
In this paper, we propose the natural inflation scenario in 
the framework of type IIB string theory on toroidal 
orientifold or orbifold and the inflaton is identified as 
the imaginary part of the complex structure moduli, Im\,$U^2$. 
The axion decay constant of inflaton is enhanced 
to the trans-Planckian field value due to the inverse of 
one-loop factor in the gauge threshold 
corrections which have a dependence 
on the complex structure moduli. 
The sections are organized as follows. 
We briefly review the gauge threshold corrections caused 
by the massive open strings between D-branes 
in the ${\cal N}=2$ sector of type II string theory in Sec.~\ref{sec:1}. 
In Sec.~\ref{sec:2}, we show the moduli stabilization procedure 
step by step and identify the lightest mode (Im\,$U^2$) 
as the inflaton. 
First, some linear combinations of dilaton and the complex structure moduli 
expect for the inflaton sector $U^2$ can be stabilized by three-form fluxes at 
the perturbative level. Second, we consider the remaining orthogonal 
linear combination of dilaton and complex structure moduli and 
K\"ahler modulus stabilization by such non-perturbative 
effects as those employed in the racetrack 
scenario~\cite{Krasnikov:1987jj} in Secs.~\ref{subsec:1}, 
\ref{subsec:2} and as that adopted in the Kachru-Kallosh-Linde-Trivedi (KKLT) 
scenario~\cite{Kachru:2003aw} in Sec.~\ref{subsec:2}. 
Then the real part of complex structure moduli Re\,$U^2$ 
can be also stabilized due to those nonvanishing superpotential 
terms at the same time. 
Finally, we extract the effective inflaton potential which is in a type of 
natural inflation with the trans-Planckian axion decay constant by 
identifying Im\,$U^2$ as the inflaton in the large complex 
structure moduli limit, ${\rm Re}\,U^2> 1$ in 
Sec.~\ref{subsec:3}. 
On the other hand, in the case of ${\rm Re}\,U^2\simeq 1$, 
we find the modulations to the original scalar potential for the natural inflation 
to be discussed in Sec.~\ref{subsec:4}. 
Sec.~\ref{sec:con} is devoted to the conclusion. 
We show the mass-squared matrices of moduli in Appendix~\ref{app:can}.

\section{Moduli-dependent threshold corrections 
in Type II string theory}
\label{sec:1}
We briefly review the one-loop stringy threshold 
corrections to the gauge couplings on D-branes 
in the framework of type II string theory. 
(For more details, see 
Refs.~\cite{Lust:2003ky,Blumenhagen:2006ci}, 
and references therein.) 
The running gauge coupling for scale $\mu$ 
below the string scale $M_s$ is written by
\begin{align}
\frac{1}{g_a^2(\mu)} =\frac{1}{g_a^2} 
+b_a\ln \left(\frac{M_s^2}{\mu^2}\right) +\frac{\Delta_a}{16\pi^2},
\end{align}
where $g_a$ is the $4$D gauge coupling at the string 
scale $M_s$, $b_a$ is the beta-function coefficient 
of the gauge group $G_a$ and $\Delta_a$ represents 
the correction from stringy massive modes at the 
one-loop level. 
In type II string theory, in general, the charged open strings 
between two stacks of D-branes or O-planes contribute to the 
gauge couplings on D-branes as the threshold corrections 
$\Delta_a$ which are mostly moduli-dependent~\cite{Dixon:1990pc}. 

In the case of type IIA string theory on 
toroidal orientifold or orbifold with O-planes and D$6$-branes 
wrapping on a supersymmetric three-cycle (special Lagrangian 
submanifold) of the internal tori, the gauge threshold corrections 
are explicitly computed by an exact CFT method 
via the cylinder and M\"obius diagram~\cite{Lust:2003ky}. 
(There are similar computations in 
type IIB string theory and F-theory on the local geometric cycle 
with fractional D-branes~\cite{Conlon:2009kt,Conlon:2009qa}.) 
When we consider the T-dual picture, 
they correspond to the setup of D$3$/D$7$-branes or 
D$5$/D$9$-branes in type IIB orientifold or orbifold 
which depend on the choice of T-duality. 
For ${\cal N}=2$ SUSY sector in type IIB string 
with D$3$/D$7$-branes and O$3$/O$7$-planes, 
(which correspond to the stacks of 
parallel D$6$- and D$6^\prime$-branes or O-planes in type IIA string theory), 
one-loop gauge threshold corrections for 
the gauge theory living on D$7$-branes with the gauge 
group $G_a$ and the wrapping numbers $(p_a^k, q_a^k)$ 
on three two-tori labeled by $k=1,2,3$ are expressed as 
\begin{align}
\Delta_{a}= -\sum_c b_{ac}^{N=2} 
\left[ \ln |\eta (i\,U^k)|^4 
+\ln \left({\rm Re}\,U^k
\frac{|p_a^k+i\,q_a^k{\rm Re}\,T^k|^2}{{\rm Re}\,T^k} \right) 
-\kappa 
\right],
\label{eq:thres}
\end{align}
where $T^k$ and $U^k$ are K\"ahler and complex structure 
moduli, respectively, 
$\kappa$ is the IR regularization constant 
and $\eta$ is the Dedekind eta-function. 
The beta-function coefficients $b^{N=2}_{ac}$ 
represent contributions from the charged 
massive modes in open strings stretched between 
the $a$-stack of D$7$-branes and the other 
$c$-stack of D-branes, and the summation over 
$c$ implicitly extracts the all 
contributions from the other stacks of D-branes. 
Note that the imaginary part 
of $T^k$ is given by the Neveu-Schwarz field. 

As pointed out in Ref.~\cite{Akerblom:2007uc}, 
only holomorphic threshold 
corrections contribute to the gauge kinetic function 
on D$7$-branes, 
which is extracted from the first term on the right-handed 
side of Eq.~(\ref{eq:thres}),
\begin{align}
f_{a}^{\rm 1-loop}=-\frac{1}{4\pi^2} 
\sum_c b^{N=2}_{ac}\ln \left( \eta (i\,U^k)\right).
\label{eq:gaugekin}
\end{align} 
Especially, in the large complex moduli limit, 
the logarithmic factor in Eq.~(\ref{eq:gaugekin}) 
behaves as 
\begin{align}
\ln \eta (i\,U^k) \rightarrow 
-\frac{\pi}{12}U^k,
\label{eq:Dedekind}
\end{align}
due to the asymptotic form of the Dedekind 
eta-function. In this limit, the gauge kinetic function 
on D$7$-branes receives the 
following threshold correction,
\begin{align}
f_{a} \simeq \sum_i \frac{T^i}{4\pi} 
+\sum_j \frac{b^j}{48\pi} U^j,
\label{eq:threscor}
\end{align}
where the summations of K\"ahler and 
complex structure moduli are only performed 
over the cycle wrapped by 
the D$7$-branes and $b^j$ represents the 
contribution from the massive open-string 
modes. 
Here we consider the D$7$-branes 
, otherwise the dilaton dependence also appears 
in the gauge kinetic function depending on the 
two-form fluxes, because such fluxes are irrelevant 
in our scenario of moduli stabilization and inflation. 
The case with a more general form of 
gauge kinetic function in terms of the 
Dedekind eta-function in Eq.~(\ref{eq:thres}) are 
discussed later. 

In the following, we propose the moduli stabilization 
and inflation scenario in the framework of type IIB 
string theory on toroidal orientifold or orbifold such 
as $T^2/Z_2$ or $T^2/(Z_2\times Z_2)$ with D-branes.

\section{Natural inflation in Type IIB string theory on 
toroidal orientifold or orbifold}
\label{sec:2}
In this section, we propose the natural inflation 
in the framework of type IIB string theory on 
toroidal orientifold or orbifold such 
as $T^2/Z_2$ or $T^2/(Z_2\times Z_2)$ with D-branes. 
The inflaton is considered as the axion 
paired with one of the complex structure moduli 
into ${\cal N}=1$ SUSY multiplet 
and the axion decay constant is enhanced to 
trans-Planckian value due to the inverse of 
a loop-factor accompanying the one-loop 
corrections to the gauge kinetic function 
which makes it possible to realize a successful natural 
inflation, as shown later. 

As pointed out in Ref.~\cite{Giddings:2001yu}, 
in the type IIB string theory 
(unlike the heterotic string theory) on 
Calabi-Yau (CY) three-fold, 
three-form fluxes induce the superpotential $W_{\rm flux}$ 
which depends on the dilaton $S$ and complex 
structure moduli $U^k$ as
\begin{align}
W_{\rm flux}=\int_{\rm CY} G_3 \wedge \Omega,
\label{eq:fsp}
\end{align}
where $\Omega$ is the holomorphic three-form of 
the CY manifold and $G_3=F_3-i\,SH_3$ is the three-form 
flux determined by Ramond-Ramond (RR) three-form 
flux $F_3$ and Neveu-Schwarz (NS) three-form flux $H_3$. 
Such flux-induced superpotential can stabilize the dilaton 
and all complex structure moduli at the perturbative 
level~\cite{Giddings:2001yu}. 

In order to show the essential idea of our scenario, 
as mentioned above, we consider the type IIB string on the 
simple toroidal orientifold or orbifold such as $T^2/Z_2$ or $T^2/(Z_2\times Z_2)$ 
whose moduli are characterized by dilaton $S$, 
three complex structure moduli $U^1$, $U^2$, $U^3$ 
and single overall K\"ahler modulus 
$T$.\footnote{It is straightforward to extend 
our stabilization mechanism to the case with three K\"ahler moduli.}
In order to obtain the desired inflation potential, 
we follow a similar step to the KKLT scenario~\cite{Kachru:2003aw} 
for stabilizing all the moduli other than ${\rm Im}\,U^2$ which is 
identified as the inflaton field.

\subsection{Moduli stabilization with three-form fluxes}
\label{subsec:1}
First, let us focus on the stabilization of the dilaton and 
complex structure moduli by employing the three-form 
flux. 
We consider the following K\"ahler potential and superpotential 
of $S$, $U^1$, $U^2$ and $U^3$ 
in the framework of $4$D ${\cal N}=1$ supergravity,
\begin{align}
K&=-\ln( S+\bar{S})-\sum_{i=1}^3\ln (U^i+\bar{U}^i),
\nonumber\\
W_{\rm flux}&=w_1+i w_2\,(U^1-U^2) +iw_3\,U^3+iw_4\,S 
+w_5U^3\,(U^1-U^2) +w_6S\,U^3+w_7S\,(U^1-U^2)
\nonumber\\ 
&+iw_8SU^3(U^1-U^2) , 
\label{eq:KW}
\end{align}
in the Planck unit, $M_{\rm Pl}=1$, where all the 
dimensionful quantities are measured by the reduced 
Plank mass\footnote{Here and hereafter, 
we adopt the Planck unit.} 
$M_{\rm Pl}=2.4\times 10^{18} {\rm GeV}$ 
and then the coefficients $w_m$ ($m=1,2,\ldots,8$) 
are integers determined by the RR- and 
NS-fluxes.\footnote{We choose the certain ansatz of three-form flux 
that yields the superpotential terms in Eq.~(\ref{eq:KW}) through Eq.~(\ref{eq:fsp}) 
in order to realize the moduli inflation as discussed later.} 

To brighten the outlook for analyzing the stabilization of dilaton 
$S$ and complex structure moduli $U^1$, $U^2$ and $U^3$, 
we redefine one of the latter as
\begin{align}
U^4=U^1-U^2.
\end{align}
In the field base $S$, $U^2$, $U^3$ and $U^4$, 
the superpotential and K\"ahler potential are rewritten as 
\begin{align}
K&= -\ln( S+\bar{S})-\ln (U^2+\bar{U}^2)-\ln (U^3+\bar{U}^3)
-\ln (U^4+\bar{U}^4+U^2+\bar{U}^2),
\nonumber\\
W_{\rm flux}&=w_1+iw_2\,U^4 +iw_3\,U^3+iw_4\,S 
+w_5U^3\,U^4 +w_6S\,U^3+w_7S\,U^4+iw_8SU^3U^4.
\label{eq:KW2}
\end{align}
Then the vacuum expectation values of dilaton and complex 
structure moduli are determined by the supersymmetric 
condition, 
\begin{align}
D_IW=0,
\label{eq:susy}
\end{align}
where $D_{I}W=W_{I}+K_{I}W$, with 
$W_{I}=\partial W/\partial \Phi^I$ and 
$K_{I}=\partial K/\partial \Phi^I$, is the covariant 
derivative with respect to the moduli fields $\Phi^I$, $\Phi^I=S,U^2,U^3$ 
and $U^4$. 

The above stabilization condition~(\ref{eq:susy}) can be satisfied 
by
\begin{align}
W_S=W_{U^3}=W_{U^4}=W=0.
\label{eq:1stab}
\end{align}
For simplicity and concreteness, we further restrict the RR- and 
NS-fluxes to those satisfying
\begin{align}
w_1=w_2\,w_6,\,\,\,
w_3=-w_5\,w_6,\,\,\,
w_4=-w_6\,w_7,\,\,\,
w_8=1,
\label{eq:fluxcond}
\end{align}
with which the expectation value of $S$, $U^3$ and $U^4$ are given by 
\begin{align}
{\rm Re}\,U^3{\rm Re}\,S=-(w_2+w_5\,w_7),\,\,\,
{\rm Re}U^4=0,\,\,\,
{\rm Im}\,U^3=w_7,\,\,\,{\rm Im}\,U^4=w_6,\,\,\,
{\rm Im}\,S=w_5,
\label{eq:u3u4s}
\end{align}
at the minimum given by 
Eq.~(\ref{eq:1stab}). 
Thus $U^4$ and the linear combination of $S$ and $U^3$ can be stabilized at the supersymmetric Minkowski minimum and their mass-squared matrices are 
found as 
\begin{align}
&m_{S}^2=\nonumber\\
&
\begin{pmatrix}
K^{U^3\bar{U}^3}|W_{U^3U^4}|^2 +K^{S\bar{S}}|W_{SU^4}|^2&
0 
& 0 
\\
0 &
K^{U^4\bar{U}^4}|W_{U^3U^4}|^2 & K^{U^4\bar{U}^4}W_{U^3U^4}\bar{W}_{\bar{S}\bar{U}^4} 
\\
0 & 
K^{U^4\bar{U}^4}\bar{W}_{\bar{U}^3\bar{U}^4}W_{SU^4} 
& K^{U^4\bar{U}^4}|W_{SU^4}|^2 
\label{app:mass1}
\end{pmatrix}
,
\end{align}
in the field basis $(U^4, U^3, S)$, which has rank $2$ 
with some appropriate choices of the integers $w_m$.~\footnote{
In Ref.~\cite{Blumenhagen:2014nba}, it is shown that when 
one leaves one modulus massless at the supersymmetry 
breaking minimum, there appears another massless moduli.} 
Note that since there is a K\"ahler mixing between $U^2$ 
and $U^4$ as can be seen from Eq.~(\ref{eq:KW2}), 
we have to canonically normalize them as 
summarized in the Appendix~\ref{app:can}. 
The K\"ahler mixing between 
$U^2$ and $U^4$ are the essential ingredients to realize the 
supersymmetric Minkowski minimum in the physical domain of 
the moduli space. 

We remark that, with the above choice of RR- and NS-fluxes, 
the tadpole cancellation condition may not occur among 
themselves, however, our moduli stabilization and natural inflation 
scenario would not depend on the detail structure of tadpole 
condition. 

\subsection{Light moduli stabilization with non-perturbative effects}
\label{subsec:2}
Next, we discuss the stabilization of remnant complex structure 
modulus $U^2$, the linear combination of $S$ and $U^3$, 
K\"ahler modulus $T$ below the mass scale of 
the stabilized complex structure moduli $U^4$. 
As mentioned above, for simplicity, we focus on the 
case with a single overall K\"ahler modulus $T$ and then its 
K\"ahler potential is expressed as 
\begin{align}
K=-3\ln ( T+\bar{T}),
\label{eq:KT}
\end{align}
in the large volume limit. As a source of stabilizing the K\"ahler 
modulus $T$, dilaton $S$ and ${\rm Re}\,U^2$, 
we assume the non-perturbative effects such as the 
gaugino condensation on D$7$-branes and  D$3$-brane,
\begin{align}
W_{\rm non}=A(U)e^{-\frac{8\pi^2 f_1}{N_1}}
-B(U)\,e^{-\frac{8\pi^2 f_2}{N_2}} 
+C(U)e^{-\frac{8\pi^2 f_3}{N_3}}
-D(U)\,e^{-\frac{8\pi^2 f_4}{N_4}},
\label{eq:gauginoW}
\end{align}
where $f_1$ and $f_2$ denote the gauge kinetic functions 
of pure $SU(N_1)\times SU(N_2)$ gauge theories on D$7$-branes,
\begin{align}
&f_1=f_2=\frac{T}{4\pi}+\frac{b^2U^2}{48\pi},
\label{eq:gkf}
\end{align}
where we assume that both of them receive 
the same threshold corrections depending on the 
complex structure modulus $U^2$ determined by 
the size of $b^2$ given by Eq.~(\ref{eq:threscor}). 
$f_3$ and $f_4$ denote the gauge kinetic functions 
of pure $SU(N_3)\times SU(N_4)$ gauge theories on D$3$-branes 
at the orbifold fixed points,
\begin{align}
&f_3=f_4=\frac{S}{4\pi}.
\label{eq:gkf}
\end{align}
$A(U)$, $B(U)$, $C(U)$ and $D(U)$ are 
functions of only the heavy complex structure 
moduli stabilized by the flux-induced superpotential~(\ref{eq:fsp}). 
Thus we can treat these functions $A(U)$ and $B(U)$ as constants, 
neglecting the fluctuations of these heavy moduli around the stabilized value. 

In the same way as employed in the Sec.~\ref{subsec:1}, we redefine 
the K\"ahler modulus as 
\begin{align}
\tilde{T}=T+\frac{b^2}{12}U^2.
\end{align}
Then the stabilization of the K\"ahler modulus $\tilde{T}$ 
and dilaton $S$ can be achieved by 
two gaugino-condensation terms 
in the same way as the racetrack scenario~\cite{Krasnikov:1987jj}, 
i.e., 
\begin{align}
D_{\tilde{T}}W_{\rm non}=(W_{\rm non})_{\tilde{T}}+K_{\tilde{T}}W_{\rm non}=0, 
\nonumber\\
D_{S}W_{\rm non}=(W_{\rm non})_{S}+K_{S}W_{\rm non}=0, 
\end{align}
which leads to the following value of the K\"ahler moduli and dilaton 
at the racetrack minimum,
\begin{align}
\langle \tilde{T} \rangle \simeq \frac{N_1N_2}{2\pi (N_2-N_1)}\ln 
\frac{N_2\,A}{N_1\,B},\,\,\,\,\,
\langle S \rangle \simeq \frac{N_3N_4}{2\pi (N_4-N_3)}\ln 
\frac{N_4\,C}{N_3\,D},
\label{eq:t}
\end{align}
where the explicit values of parameters are 
explored by evaluating the cosmological observables. 
The above racetrack minimum can be realized due to the following relation, $\langle W_{\rm non}\rangle \ll \langle 
(W_{\rm non})_{\tilde{T}} \rangle, \langle 
(W_{\rm non})_{S} \rangle$ which is satisfied 
in our parameter regions as shown later. 
As mentioned in Sec.~\ref{subsec:1}, the linear combination of $S$ and $U^3$ is already stabilized by the flux induced superpotential. Here the remaining  orthogonal combination can be stabilized by this 
racetrack superpotential for $S$. (The stabilization 
point of ${\rm Im}\,S$ is the same as that given by 
Eq.~(\ref{eq:u3u4s}) when we choose $w_5=0$ in the 
superpotential~(\ref{eq:KW2}).)

In the following, let us discuss the stabilization of the remnant 
complex structure modulus $U^2$. 
Since the K\"ahler modulus is stabilized at the minimum 
$\langle W\rangle \neq 0$, 
the real part of $U^2$ is stabilized by the K\"ahler potential,
\begin{align}
K=-\ln (U^2+\bar{U}^2) -\ln (U^4+\bar{U}^4 +U^2+\bar{U}^2)
-3\ln ( \tilde{T}+\bar{\tilde{T}}-\frac{b^2}{12}(U^2+\bar{U}^2)),
\label{eq:K3}
\end{align}
under the following condition:
\begin{align}
K_{U^2}=-\frac{1}{U^2+\bar{U}^2}-\frac{1}{U^4+\bar{U}^4+U^2+\bar{U}^2}
+\frac{b^2}{4}\frac{1}{\tilde{T}+\bar{\tilde{T}}-\frac{b^2}{12}(U^2+\bar{U}^2)}=0,
\label{eq:u2}
\end{align}
which determines the expectation value of Re\,$U^2$ as 
\begin{align}
{\rm Re}\,U^2=\frac{24\langle{\rm Re}\,\tilde{T}\rangle}{5b^2},
\end{align}
satisfying the extremal condition $V_{U^2}=\partial V/\partial U^2=0$ 
and we employed $\langle {\rm Re}\,U^4\rangle=0$. 
We again remark that $U^4$ and the linear combination of 
$U^3$ and $S$ are 
fixed at a high-scale by the condition 
$D_{U^3}W=D_{U^4}W=D_{S}W=0$. 
Therefore, if the gaugino condensation scale is much smaller 
than the mass scale of $U^4$ and the linear combination of 
$U^3$ and $S$, 
their deviations from the minimum given by 
Eq.~(\ref{eq:u3u4s}) are sufficiently small, and 
we can replace the heavy moduli 
$U^4$, the linear combination of $U^3$ and 
$S$ by their expectation values~(\ref{eq:u3u4s}) 
in evaluating the stabilization of light moduli $\tilde{T}$, $S$ and 
Re\,$U^2$. 
 
To confirm the stabilization of Re\,$U^2$, we have to check that 
the rank of the full mass-squared matrices for $U^2$, $U^3$, $U^4$, $S$ 
and $\tilde{T}$. 
The explicit form of them and the canonical normalization of all moduli 
are summarized in Appendix~\ref{app:can}. 
From the mass matrices shown in Eq.~(\ref{app:mass}), we find the squared mass 
of Re\,$U^2$ is positive, if the mass scales of $U^4$ and the 
linear combination of $U^3$ and $S$ are 
much heavier than the gaugino condensation scale determined by 
the superpotential~(\ref{eq:gauginoW}), that is consistent with the 
above argument. 

In the above analysis, the vacuum energy is negative at the minimum 
$D_I W=0$ for $I=U^2,U^3,U^4,S$ and $\tilde{T}$. Therefore we assume the existence 
of some uplifting sector with which the total scalar potential $V$ 
is in KKLT-type~\cite{Kachru:2003aw},
\begin{align}
V=V_F +V_{\rm up},
\end{align}
where $V_F$ is written by the usual ${\cal N}=1$ 
supergravity formula,
\begin{align}
V_F=e^K\left (K^{I\bar{J}}D_I WD_{\bar{J}}\bar{W}-3|W|^2\right),
\label{eq:VF}
\end{align}
with $K^{I\bar{J}}$ is the inverse of the K\"ahler metric 
$K_{I\bar{J}}=\partial^2 K/\partial \Phi^I\partial\bar{\Phi}^{\bar{J}}$ 
for $\Phi^I=S,\tilde{T},U^2,U^3$ and $U^4$. 
The uplifting potential $V_{\rm up}$ may come from 
anti-D$3$-branes~\cite{Kachru:2003aw} and/or nonvanishing 
F-terms in some dynamical SUSY breaking 
sector~\cite{Dudas:2006gr,Abe:2006xp,Kallosh:2006dv,Abe:2007yb}, 
etc.. 
In the next section, we show the inflaton potential by identifying 
the light axion Im\,$U^2$ as the inflaton.

Finally, we comment on the stabilization of the 
K\"ahler modulus $T$. 
In the above analysis, the K\"ahler modulus is 
stabilized in the same way as the racetrack 
scenario~\cite{Krasnikov:1987jj}. 
In the case of the KKLT scenario~\cite{Kachru:2003aw} 
which is achieved, instead of Eq.~(\ref{eq:gauginoW}), 
by a single gaugino-condensation term and a tiny constant 
value of the nonvanishing flux-induced superpotential, 
we can also derive the similar inflaton potential with 
a trans-Planckian axion decay constant 
as seen in the next section. 
This is because our inflaton potential does not depend 
on the stabilization of $T$. 
However, in the latter case, we need to tune the 
RR- and NS-flux to obtain the tiny expectation value of 
superpotential $\langle W\rangle \simeq 10^{-2}$ 
in order to realize the large volume limit required to 
ensure the form of K\"ahler potential shown in Eq.~(\ref{eq:KT}). 
When the three-form fluxes are turned on, we may have 
to consider a more general geometry than CY (locally) 
warped due to the energies of these fluxes as well as some 
sources for the tadpole cancellation~\cite{Giddings:2001yu}. 
Thus we further assume that the possible backreactions from 
the three-form fluxes are negligible in the relevant sector to 
our scenario of moduli stabilization and inflation. 

So far, we focus on the single overall K\"ahler modulus $T$. 
The other K\"ahler moduli $T_i$ $i=1,2,\cdots$, are also 
stabilized by the non-perturbative effects such as the 
gaugino condensation on D$7$-branes 
irrelevant to the cycle associated with the modulus $T$, 
\begin{align}
W=\sum_{i}A_i(U)e^{-\frac{8\pi^2 f_1^{(i)}}{M_1^{(i)}}}
-B_i(U)\,e^{-\frac{8\pi^2 f_2^{(i)}}{M_2^{(i)}}},
\end{align}
at the racetrack minimum, 
where $A_i(U)$, $B_i(U)$ are the functions of only the 
heavy complex structure moduli by the flux-induced 
superpotential~(\ref{eq:fsp}). 
$f_1^{(i)}$ and $f_2^{(i)}$ denote the gauge kinetic functions of 
$SU(M_1^{(i)})\times SU(M_2^{(i)})$ gauge theories, e.g., 
$f_1^{(i)}=a_1T_i$ and $f_2^{(i)}=a_2T_i$ with $a_1$, $a_2$ 
are constants. 
It is then assumed these gaugino condenstion scales are 
much heavier than the those for the modulus $T$. 

\subsection{Natural inflation without modulations}
\label{subsec:3}
Now we are ready to write down the inflaton potential. 
The effective scalar potential for Im\,$U^2$ 
is generated from another $SU(L)$ gaugino-condensation term,
\begin{align}
W \supset E(\langle U\rangle) 
e^{-\frac{2\pi}{L}\langle T\rangle 
-\frac{b\pi}{6L}\langle{\rm Re}\,U^2\rangle 
-i\frac{b\pi}{6L}{\rm Im}\,U^2},
\label{eq:Wu2}
\end{align}
where we assume the gauge coupling on $SU(L)$ gauge theory 
receives the threshold corrections which have $U^2$-dependence. 
The factor $E(\langle U\rangle)$ denotes possible threshold 
corrections from the heavy complex structure moduli, $U^3$ and $U^4$, 
as in the previous step. 
We assume again that all the other moduli are strictly fixed 
at their minimum given by Eqs.~(\ref{eq:u3u4s}) and~(\ref{eq:t}) 
obtaining heavy masses 
and the fluctuations around their vacuum expectation values 
are neglected in the effective potential for Im\,$U^2$. 
Such a situation can be realized if the rank of the $SU(L)$, 
$SU(N_i)$ ($i=1,2,3,4$) gauge theories are chosen 
as $L < N_i$ with $i=1,2,3,4$. 
In this case, 
Im\,$U^2$ is lighter enough than all the other moduli 
those receive much heavier masses from the 
high-scale gaugino-condensation terms~(\ref{eq:gauginoW}) 
and the flux-induced superpotential~(\ref{eq:fsp}), respectively. 

After all, the effective scalar potential for Im\,$U^2$ 
is generated from $V_F$ in Eq.~(\ref{eq:VF}), which is 
written as
\begin{align}
V_{\rm eff} =\Lambda_1 \left(1-{\rm cos} 
\left(\lambda_1 \phi\right) \right),
\label{eq:infpo}
\end{align}
where 
$\Lambda_1\simeq 6e^{\langle K\rangle}
\langle W_{\rm non}\rangle E(\langle U\rangle) 
e^{-\frac{2\pi}{L}\langle T\rangle -\frac{b\pi}{6L}{\rm Re}
\langle U^2\rangle}$, $\lambda_1=b\pi/6d\,L$ and 
$\phi=d\,{\rm Im}\,U^2$ is the canonically normalized axion 
field. The normalization factor 
$d\simeq 1/\langle{\rm Re}\,U^2\rangle$ 
is determined by the 
canonical normalization of relevant complex structure 
moduli which is explicitly shown in Appendix~\ref{app:can}. 
Even though $U^2$, $U^4$ and $\tilde{T}$ have a kinetic-mixing from 
the structure of K\"ahler potential~(\ref{eq:K3}) as mentioned 
in Sec.~\ref{subsec:2}, the effects from the mixing between 
$\phi$ and ${\rm Im}\,U^4$, Im\,$\tilde{T}$ is negligible on the inflation mechanism 
discussed in the following. This is because Im\,$U^4$ and Im\,$\tilde{T}$ 
are heavier enough than Im\,$U^2$ and already decoupled 
from the inflaton dynamics. 

When we identify the inflaton as $\phi$, the 
axion potential is considered as the type of natural inflation. 
As seen in the scalar potential~(\ref{eq:infpo}), 
the axion decay constant is enhanced by the inverse of 
one-loop factor and is determined by 
the ratio $b/L$ and the vacuum expectation value, 
$\langle{\rm Re}\,U^2\rangle$. 
Since the trans-Planckian axion decay constant is 
required in order to explain the cosmological 
observations reported by Planck data~\cite{Ade:2015lrj}, 
the ratio 
$b/L$ and $\langle{\rm Re}\,U^2\rangle$ have to be 
properly chosen. 
Note that the beta-function coefficient $b$ in 
${\cal N}=2$ sector is not related with 
the sector of $SU(L)$ gauge theory. 

To evaluate the cosmological observables, 
we define the slow-roll parameters for the inflaton $\phi$, 
\begin{align}
&\epsilon =\frac{1}{2}\left( \frac{\partial_{\phi}V_{\rm eff}}{V_{\rm eff}}\right)^2 
=\frac{(\lambda_1)^2}{2} 
\frac{1-{\rm cos}^2 (\lambda_1\,\phi)}
{\left( 1-{\rm cos} (\lambda_1\,\phi)
\right)^2}, \nonumber\\
&\eta =\frac{\partial_{\phi}\partial_{\phi}V_{\rm eff}}{V_{\rm eff}}
=(\lambda_1)^2 
\frac{{\rm cos} (\lambda_1\,\phi)}
{1-{\rm cos} (\lambda_1\,\phi)},
\nonumber\\
&\xi =
\frac{\partial_{\phi}V_{\rm eff} 
\partial_{\phi}\partial_{\phi}\partial_{\phi}V_{\rm eff}}{V_{\rm eff}^2}
=-(\lambda_1)^4 
\frac{1-{\rm cos}^2 (\lambda_1\,\phi)}
{\left( 1-{\rm cos} (\lambda_1\,\phi)
\right)^2}, 
\end{align}
and then the cosmological observables such as 
the power spectrum of the scalar density perturbation 
$P_\zeta$, the spectral index $n_s$, its running $dn_s/d\ln k$ 
and the tensor-to-scalar ratio $r$ are 
written as 
\begin{align}
P_{\zeta}=\frac{V}{24\pi^2\,\epsilon},\,\,\,\,
n_s=1+2\eta -6\epsilon,\,\,\,\,
r=16\epsilon,\,\,\,\,
\frac{d\,n_s}{d\,\ln k}=16\epsilon\eta 
-24\epsilon^2 -2\xi,
\end{align}
by employing the slow-roll approximation 
at the leading order. The e-folding number 
is also evaluated as 
\begin{align}
N_e=\int_{\phi_{\rm end}}^{\phi} 
\frac{V_{\rm eff}}{\partial_{\phi}V_{\rm eff}} d\phi, 
\end{align}
where $\phi_{\rm end}$ denotes the 
field value at the end of inflation with 
which the slow-roll condition is violated, 
max$\{\epsilon, \eta, \xi \}=1$. 

In order to explain the power spectrum of 
the scalar density perturbation, 
$P_{\zeta}\simeq 2.2 \times 10^{-9}$ reported 
by Planck~\cite{Ade:2015lrj}, 
we set the parameters in the 
superpotential given by Eqs.~(\ref{eq:KW2}), 
(\ref{eq:t}), (\ref{eq:Wu2}) and the K\"ahler 
potential given by Eq.~(\ref{eq:K3}), 
\begin{align}
&w_5=0,\,\,\,w_6=1,\,\,\,w_2=-8,\,\,\,w_7=-3,\,\,\,
N_1=N_3=12,\,\,\, N_2=N_4=20,\,\,\, L=10,\,\,\,
b=1,
\nonumber\\
&b^2=12,\,\,A=-8,\,\,B=-3,\,\,C=9,\,\,D=3,\,\,E=\frac{1}{12},
\label{eq:para1}
\end{align}
and the other parameters in Eq.~(\ref{eq:KW2}) 
are fixed such that the conditions given by 
Eq.~(\ref{eq:fluxcond}) are satisfied, those lead to the 
vacuum expectation values of moduli, 
\begin{align}
&\langle {\rm Re}\,U^1\rangle \simeq 
\langle {\rm Re}\,U^2\rangle \simeq 2.8,\,\,
\langle{\rm Re}\,U^3 \rangle \simeq 1,\,\,
\langle{\rm Im}\,U^1 \rangle \simeq 1,\,\,
\langle{\rm Im}\,U^2 \rangle \simeq 0,\,\,
\langle{\rm Im}\,U^3 \rangle \simeq -3,
\nonumber\\
&\langle {\rm Re}\,S\rangle \simeq 7.7,\,\,
\langle {\rm Im}\,S\rangle \simeq 0,\,\,
 \langle \hat{T}\rangle\simeq 7.1. 
\label{eq:mvev}
\end{align}
 
With the above set of parameters, the above 
cosmological observables and the e-folding number 
are evaluated as 
\begin{align}
n_s\simeq 0.963,\,\,\, r\simeq 0.06,\,\,\,
dn_s/d\ln k\simeq -8\times 10^{-4},\,\,\,N_e\simeq 61,
\end{align}
which are consistent with 
WMAP, Planck data~\cite{Ade:2015lrj},
\begin{align} 
n_s= 0.9655\pm 0.0062,
\end{align}  
at the pivot scale $k_{\ast}=0.05 {\rm Mpc}^{-1}$ 
and the upper limit of $r$~\cite{Ade:2015lrj},
\begin{align} 
r<0.11,
\label{eq:obtensor}
\end{align}  
when we properly choose the initial condition 
for inflaton $\phi$. 

We remark that, in our model, 
the axion decay constant is 
enhanced by the inverse of loop factor through the stringy 
threshold corrections which are characterized by the 
Dedekind eta-function and the beta-function 
coefficients $b$ induced by the massive open-string modes 
between D-branes in Eq.~(\ref{eq:gkf}). 
Thus, we can realize several values of axion 
decay constant depending on the brane configurations, 
which means that the tensor-to-scalar 
ratio can be of ${\cal O}(0.01-0.1)$ in our framework. 

\subsection{Natural inflation with modulations}
\label{subsec:4}
The previous section shows the usual 
natural inflation with trans-Planckian decay constant, 
which is valid only when the Dedekind eta-function 
can be approximated by the leading term as shown 
in Eq.~(\ref{eq:Dedekind}) in the large field limit of complex 
structure moduli. 

In this section, we estimate the deviations from the large 
complex-structure limit, by introducing the next leading 
term in the Dedekind function, 
\begin{align}
\eta (i\,U^2) \rightarrow 
e^{-\frac{\pi}{12}U^2}\Bigl[1-e^{-2\pi U^2}-{\cal O}(e^{-4\pi U^2})],
\label{eq:Dedekind2}
\end{align}
which induces the following correction to the inflaton 
potential given by Eq.~(\ref{eq:infpo}),
\begin{align}
V_{\rm inf}=V_{\rm eff} +V_{\rm mod},
\label{eq:totinfpo}
\end{align}
where 
\begin{align}
V_{\rm mod}=\Lambda_2\,{\rm cos} 
\left(\lambda_2\phi\right),
\end{align}
with $\Lambda_2=\Lambda_1\,
\frac{2b}{L}e^{-\left(2\pi +\frac{b\,\pi}{6L}\right)
\langle{\rm Re}\,U^2\rangle}$, 
$\lambda_2=(2\pi +b\,\pi/6L)/d$. Note that 
the correction $V_{\rm mod}$ would in general yield the 
modulations~\cite{Kobayashi:2010pz,Czerny:2014wua,
Kobayashi:2014ooa} 
to the leading inflaton potential $V_{\rm eff}$ in the case of 
$\langle{\rm Re}\,U^2\rangle \simeq 1$, though it is not 
the case in the above analysis with the numerical values of 
parameters~(\ref{eq:para1}) resulting 
Eqs.~(\ref{eq:mvev})-(\ref{eq:obtensor}). 

In the following analysis, we also take care of 
the vacuum expectation value of ${\rm Re}\,U^2$ 
and the ratio $b/L$ in order to avoid the tachyonic 
scalar potential around the origin, $\phi=0$. 
Actually we can avoid the nonvanishing field value of the axion 
$\phi$ at the minimum, which would lead to the strong CP problem 
if it couples to the QCD sector, that is, 
the physical $\bar{\theta}$ term is severely 
constrained by the non-observation of electric dipole 
moment of the neutron~\cite{Harris:1999jx,Baker:2006ts}. 
The axion mass squared at the origin is described by 
\begin{align}
\partial^2_\phi V_{\rm inf} \bigl|_{\phi=0}
=(\lambda_1)^2\Lambda_1 -(\lambda_2)^2\Lambda_2,
\end{align}
and its positivity is ensured by the following condition:
\begin{align}
(\lambda_1)^2\Lambda_1 -(\lambda_2)^2\Lambda_2 >0\,
\leftrightarrow\,
\left(\frac{\pi}{6}\right)^2\frac{b}{L} 
> 2\left(2\pi+\frac{\pi\,b}{6L}\right)^2 e^{-2\pi 
\langle{\rm Re}\,U^2\rangle}. 
\end{align}

For general cases, it is interesting to discuss the contributions 
from the additional scalar potential~$V_{\rm mod}$ to the inflaton 
dynamics. By the inclusion of~$V_{\rm mod}$, 
the slow-roll parameters of the inflaton potential~$V_{\rm inf}$ 
for the inflaton $\phi$ are written as 
\begin{align}
&\epsilon =
\frac{\left(\lambda_1\Lambda_1\,{\rm sin}\,(\lambda_1\,\phi)
-\lambda_2\Lambda_2\,{\rm sin}\,(\lambda_2\,\phi)
\right)^2}{2\,V_{\rm inf}^2}, \nonumber\\
&\eta =
\frac{(\lambda_1)^2\Lambda_1\,{\rm cos}\,(\lambda_1\,\phi)
-(\lambda_2)^2\Lambda_2\,{\rm cos}\,(\lambda_2\,\phi)}
{V_{\rm inf}},
\nonumber\\
&\xi^2 =-
\frac{\lambda_1\Lambda_1\,{\rm sin}\,(\lambda_1\,\phi)
-\lambda_2\Lambda_2\,{\rm sin}\,(\lambda_2\,\phi)}
{V_{\rm inf}}\times \frac{(\lambda_1)^3\Lambda_1\,{\rm sin}\,(\lambda_1\,\phi)
-(\lambda_2)^3\Lambda_2\,{\rm sin}\,(\lambda_2\,\phi)}
{V_{\rm inf}},
\label{eq:slow2}
\end{align}
while the spectral index $n_s$ including the higher-order 
corrections is found as 
\begin{align}
n_s =1+2\eta -6\epsilon 
+2\biggl[ -\left( \frac{5}{3}+12C\right)\epsilon^2 
+(8C-1)\epsilon\,\eta +\frac{1}{3}\eta^2 
-\left(C-\frac{1}{3}\right)\xi^2\biggl]+\cdots,
\label{eq:nshigher}
\end{align}
where $C=-2+\ln 2 +\gamma$ with $\gamma\simeq 0.577$ is 
the Euler-Mascheroni constant and the ellipsis stands for 
more higher corrections which are given by the fourth 
derivative with respect to the inflaton.~(See Ref.~\cite{Lyth:1998xn} 
and references therein.) As discussed later, 
in models which have $\xi^2 ={\cal O}(0.01)$, 
the higher-order terms contribute to the numerical value of 
$n_s$ in Eq.~(\ref{eq:nshigher}), while 
the higher-order corrections to $P_\zeta$ 
do not give sizable effects. 
Note that our inflaton effective potential is controlled by 
$\langle{\rm Re}\,U^2\rangle$ and $b/L$ in the superpotential 
given by Eq.~(\ref{eq:KW2}). 
In the following analysis, we assume certain numerical 
values of parameters different from Eq.~(\ref{eq:para1}), 
those realize the particular value 
$\langle{\rm Re}\,U^2\rangle\simeq 1$, with which 
$V_{\rm mod}$ does affect the inflaton dynamics. 

\begin{figure}[ht]
\centering \leavevmode
\includegraphics[width=0.5\linewidth]{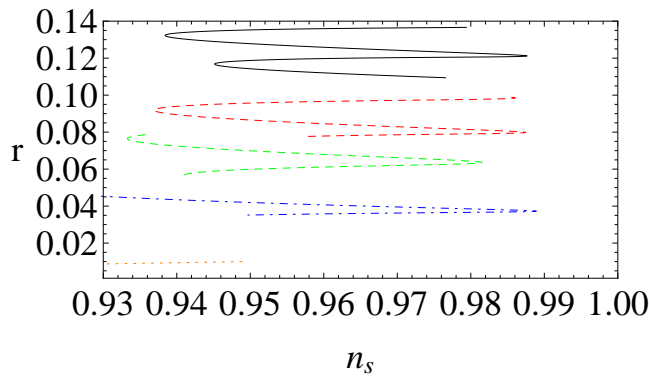}
\caption{Predictions of $(n_s, r)$ in the range of 
e-folding number, $50\leq N_e \leq 60$. 
For the universal value of $\langle{\rm Re}\,U^2\rangle=1$, 
black-solid, red-dashed, green-dashed, blue-dotdashed and 
orange-dotted lines correspond to the fixed ratios 
$b/L=1/10, 1/5, 1/4, 1/3, 1/2$, respectively.}
\label{fig:rns}
\end{figure}

\begin{figure}[h]
\begin{minipage}{0.5\hsize}
\begin{center}
\includegraphics[width=0.8\linewidth]{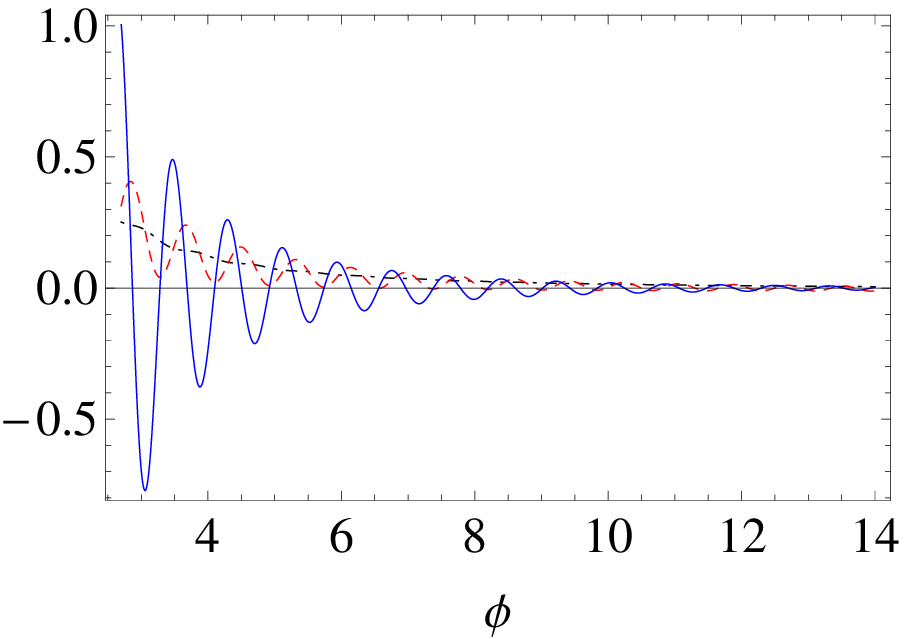}
\end{center}
\end{minipage}
\begin{minipage}{0.5\hsize}
\begin{center}
\includegraphics[width=0.8\linewidth]{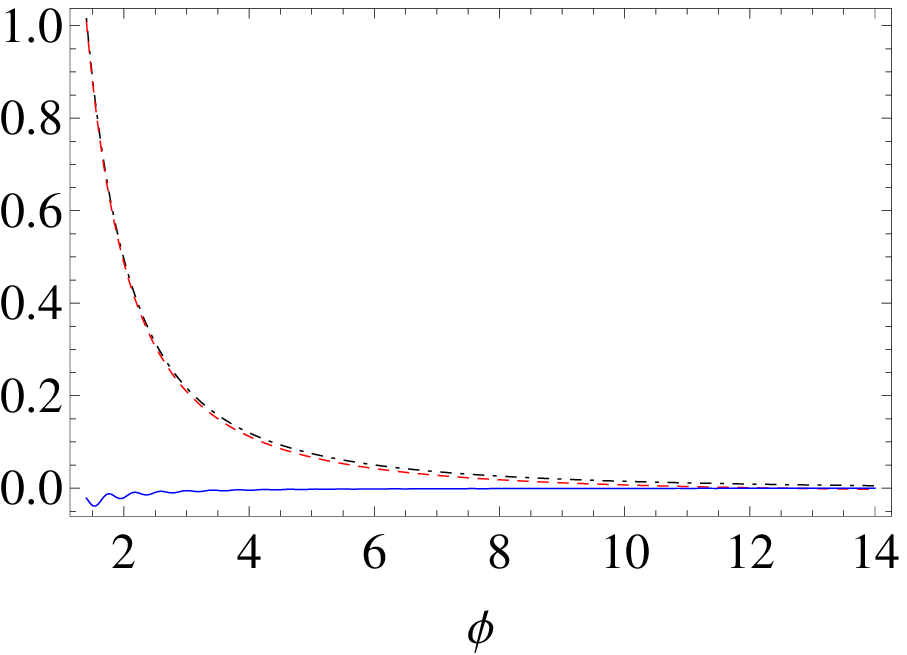}
\end{center}
\end{minipage}
\caption{The behavior of the slow-roll parameters, 
$\epsilon$, $\eta$ and $\xi^2$, which correspond to 
black-dotdashed, red-dashed and blue-solid curves, 
respectively. In the left (right) panel, we set $b/L=1/5~(1/10)$ 
and $\langle{\rm Re}\,U^2\rangle =1.2~(2.4)$.}
\label{fig:slowpara}
\end{figure}

\begin{figure}[ht]
\centering \leavevmode
\includegraphics[width=0.5\linewidth]{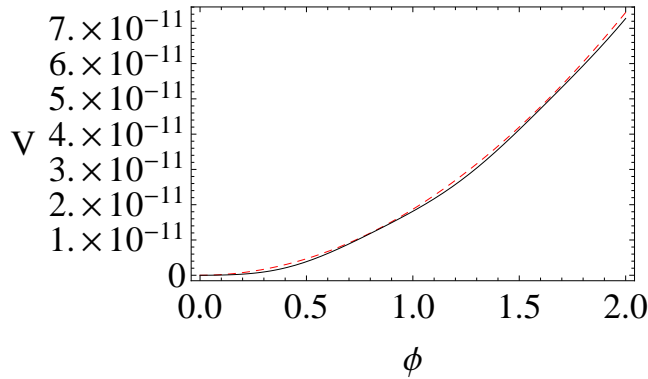}
\caption{The scalar potential $V$ versus the inflaton value 
$\phi$. Along with Fig.~\ref{fig:slowpara}, 
the black-solid curve corresponds to the scalar 
potential~(\ref{eq:totinfpo}) with modulations for the parameter $b/L=1/5$ 
and $\langle{\rm Re}\,U^2\rangle =1.2$. 
On the other hand, the red-dotted curve corresponds 
to the leading scalar potential (\ref{eq:infpo}) without modulations 
for the same parameters.}
\label{fig:vphi}
\end{figure}

Then we numerically evaluate the cosmological observables 
$r$, $n_s$, $dn_s/d\ln k$ by putting several values of 
$b/L$. The scalar density perturbation $P_\zeta$ can be 
obtained as $2.2\times 10^{-9}$ also in this case by suitably 
choosing the gaugino-condensation terms in Eq.~(\ref{eq:t}) which stabilize 
the K\"ahler modulus at the racetrack minimum. 
Fig.~\ref{fig:rns} shows the prediction of the spectral index $n_s$ 
and the tensor-to-scalar ratio $r$ in the range of 
e-folding number, $50\leq N_e \leq 60$. 
Several oscillating curves are drawn by varying $N_e$ with the 
corresponding fixed values of the ratio $b/L$ in Fig.~\ref{fig:rns}. 
This is because the slow-roll 
parameters oscillate due to the inclusion of the 
deviations from the large complex-structure limit 
as can be seen in Fig.~\ref{fig:slowpara} which 
shows the behavior of the slow-roll parameters by 
setting $b/L=1/5~(1/10)$ 
and $\langle{\rm Re}\,U^2\rangle =1.2~(2.4)$ in the 
left (right) panel. Although, in the both left and right panels 
in Fig.~\ref{fig:slowpara}, the 
leading scalar potential~$V_{\rm eff}$ has the same 
structure, the next-leading scalar potential~$V_{\rm mod}$ 
gives sizable corrections in the left rather than 
the right panel. The scalar potential with and without 
such modulations is shown in the Fig.~\ref{fig:vphi}. 
As mentioned above, the detectability of such modulations 
is governed by the expectation 
value of ${\rm Re}\,U^2$ and then the next-to-next 
leading scalar potential which comes from the expansion of 
the Dedekind functions would be important in the 
case of $\langle{\rm Re}\,U^2\rangle <1$. 
We summarize our predictions for the cosmological 
observables in Table~$1$. 


\begin{table}
\begin{center}
\begin{tabular}{|c|c|c|c|c|c|} \hline 
$b/L$ &$\langle{\rm Re}\,U^2\rangle$ & $N_e$& $n_s$ & $r$ & $dn_s/d\ln k$ 
\\ \hline
$1/10$ & $1.3$ & $50$ & $0.96$ & $0.14$ & $-0.0008$ 
\\\hline
$1/10$ & $1.3$ & $57$ & $0.96$ &$0.12$ & $-0.012$ 
\\\hline
$1/5$ & $1.2$ &  $55$ & $0.96$& $0.08$ & $-0.002$ 
\\\hline
$1/5$ & $1.2$ & $60$ & $0.96$& $0.08$ & $-0.001$ 
\\\hline
$1/4$ & $1.2$ &  $53$ & $0.96$& $0.07$ & $-0.002$ 
\\\hline
$1/4$ & $1.2$ & $58$ & $0.96$& $0.06$ & $-0.001$ 
\\\hline
$1/3$ & $1.1$ & $54$ & $0.96$& $0.04$ & $-0.002$ 
\\\hline
$1/3$ & $1.1$ & $60$ & $0.96$& $0.04$ & $-0.001$ 
\\\hline
$1/2$ & $1.1$ & $50$ & $0.95$& $0.01$ & $-0.0003$ 
\\\hline
 \end{tabular}
 \caption{The input values of $b/L$, Re\,$U^2$ and the 
output values of the e-folding number $N_e$, 
spectral index $n_s$, tensor-to-scalar 
ratio $r$ and the running of spectral index $dn_s/d\ln k$.}
\end{center}
\end{table}

Our results suggest that we can realize several 
values of the tensor-to-scalar ratio and spectral index 
independently to each other when we consider the particular 
value of complex-structure modulus, 
$\langle{\rm Re}\,U^2\rangle\simeq 1$. 
This nature is different from the 
original natural inflation model~\cite{Freese:1990rb} and is also seen in 
the multi-natural inflation scenario~\cite{Czerny:2014xja}. 
However, up to now, we do not know which amount of 
gravitational waves are observed reported by 
BICEP2 collaborations~\cite{Ade:2014xna}. 
We expect that future cosmological observations 
select more precisely certain values of cosmological observables.

In summary, in our framework of type IIB string theory on toroidal 
orientifold or orbifold, the deviation from the natural 
inflation depends on the expectation value of the real part of 
complex structure modulus, $\langle{\rm Re}\,U^2\rangle$. 
In the large field limit of complex structure moduli, 
our inflaton potential is considered as the original 
natural inflation scenario~\cite{Freese:1990rb}. 
When we construct the standard model sector on 
D$p$-branes ($p>3$), the matter fields in the standard model 
generically couple to the complex structure moduli. 
Such couplings affect the inflaton dynamics after the end of 
inflation which are related to the reheating processes. 
Thus, it is interesting to study toward such a direction 
in a future work.

\section{Conclusion}
\label{sec:con}
We proposed a mechanism for the natural inflation with 
and without modulations in the framework of type IIB string theory 
on toroidal orientifold or orbifold. 
The essential ingredient to obtain the trans-Planckian decay 
constant which is required in the natural inflation 
is the holomorphic gauge threshold corrections 
to the gauge kinetic function. 
Such threshold corrections are exactly computed 
in type II string theory on toroidal orientifold or 
orbifold by employing the CFT method 
(see Ref.~\cite{Lust:2003ky,Blumenhagen:2006ci}, 
and references therein) 
which suggests the gauge threshold corrections 
have moduli dependences. 
Note that when one of the moduli is identified 
as the inflaton, the moduli-dependent threshold 
corrections are important 
not only to discuss about the gauge coupling 
unification, but also to enhance the axion 
decay constant of the inflaton by the inverse 
of one-loop factor accompanying the correction. 

In our model, the inflaton is considered as Im\,$U^2$ 
which is the imaginary part of the complex structure 
modulus and the inflaton potential is extracted 
from the gaugino-condensation term whose 
gauge coupling receives the complex structure 
moduli-dependent terms characterized by the 
Dedekind function. 
We presented that in the large complex-structure 
limit, $\langle {\rm Re}\,U^2\rangle > 1$, 
the Dedekind function is approximated as the single 
exponential term and then the inflaton potential is 
close to that of the natural inflation which is consistent 
with cosmological observations such as WMAP, 
Planck~\cite{Planck:2013jfk,Ade:2015lrj} 
and the joint analysis of BICEP2, Keck Array and 
Planck~\cite{Ade:2015tva}. 
On the other hand, in the regime with 
$\langle {\rm Re}\,U^2\rangle \simeq 1$, 
we have to take account of the explicit Dedekind 
function, which leads to the modulations to 
the original natural inflation~\cite{Freese:1990rb}. 
The modulations give a sizable modification to 
the predictions~\cite{Kobayashi:2010pz,
Czerny:2014wua,Kobayashi:2014ooa} 
of the original natural inflation 
in the same way as the multi-natural inflation 
scenario~\cite{Czerny:2014xja}. 
The natural inflation with modulations predicts 
the different predictions unlike the original 
natural inflation without modulations. In fact, 
we can achieve the small and large tensor-to-scalar 
ratio without changing the value of spectral 
index so much. Thus such natural inflation 
with modulations can be tested in the near 
future experiments. 
 
In both inflation scenarios, we stabilize the 
complex structure moduli except for the 
inflaton sector by employing three-form fluxes 
in the usual manner. The dilaton $S$, 
overall K\"ahler modulus $T$ 
and Re\,$U^2$ are stabilized at the racetrack (KKLT) 
minimum by double (single) gaugino-condensation 
terms (term) above the inflation scale. 
In general, although it seems to be difficult to obtain 
the mass difference between Re\,$U^2$ and 
Im\,$U^2$, it can be achieved by the K\"ahler mixing 
between $U^2$ and the other moduli 
in our model. 

We have not discussed the reheating process. 
When we construct the standard model sector on 
D$p$-branes ($p>3$), the matter fields in the standard model 
generically couple to the complex structure modulus (inflaton). 
Since such couplings affect the inflaton dynamics after the end of 
inflation, it is interesting to study in such a direction for the future work. 

It is also interesting to extend our set-up to more 
general Calabi-Yau manifold. However, the one-loop 
threshold corrections are unknown on D-branes which 
wrap the internal cycles of Calabi-Yau manifold. Thus 
it is beyond our scope.

\subsection*{Acknowledgement}
H.~O. would like to thank Kiwoon~Choi and Tetsutaro~Higaki 
for useful discussions and comments. 
H.~A. was supported in
part by the Grant-in-Aid for Scientific Research No. 25800158 from the
Ministry of Education,
Culture, Sports, Science and Technology (MEXT) in Japan. T.~K. was
supported in part by
the Grant-in-Aid for Scientific Research No. 25400252 from the MEXT in
Japan.
H.~O. was supported in part by a Grant-in-Aid for JSPS Fellows 
No. 26-7296.

\appendix
\section{The canonical normalization and 
mass-squared matrices}
\label{app:can}
In this appendix, we show the canonical normalization and 
the mass-squared matrices of all moduli given by the 
K\"ahler potential~(\ref{eq:KW2}), (\ref{eq:KT}) 
and the superpotential~(\ref{eq:KW2}), (\ref{eq:gauginoW}). 

The K\"ahler metric generated by the 
K\"ahler potential~(\ref{eq:KW2}), (\ref{eq:KT}) is 
given by 
\begin{align}
K_{I\bar{J}}=
\begin{pmatrix}
K_{U^4\bar{U}^4} & K_{U^4\bar{U}^2} &0 &0 &0 \\
K_{U^2\bar{U}^4} & K_{U^2\bar{U}^2} &K_{U^2\bar{\tilde{T}}} &0 &0 \\
0 &K_{\tilde{T}\bar{U}^2} &K_{\tilde{T}\bar{\tilde{T}}} & 0 &0 \\
0 &0 &0 &K_{U^3\bar{U}^3} &0 \\
0 &0 &0 & 0 &K_{S\bar{S}} \\
\end{pmatrix}
,
\end{align} 
where 
\begin{align}
&K_{U^2\bar{U}^2}=\frac{1}{(U^2+\bar{U}^2)^2}+
\frac{1}{(U^4+\bar{U}^4+U^2+\bar{U}^2)^2}+
\frac{3c_2^2}{(\tilde{T}+\bar{\tilde{T}}-c_2(U^2+\bar{U}^2))^2}
=\frac{10}{3}\frac{1}{(U^2+\bar{U}^2)^2},
\nonumber\\
&K_{U^2\bar{U}^4}=K_{U^4\bar{U}^2} =K_{U^4\bar{U}^4}=\frac{1}{(U^2+\bar{U}^2)^2},
\,\,\,\,
K_{U^2\bar{\tilde{T}}}=K_{\tilde{T}\bar{U}^2} =-\frac{4}{3c_2}\frac{1}{(U^2+\bar{U}^2)^2},\nonumber\\
&K_{U^3\bar{U}^3}=\frac{1}{(U^3+\bar{U}^3)^2},\,\,\,
K_{S\bar{S}}=\frac{1}{(S+\bar{S})^2},\,\,\,
K_{T\bar{T}}=\frac{3}{(\tilde{T}+\bar{\tilde{T}}-c_2(U^2+\bar{U}^2))^2}
=\frac{4}{3c_2^2}\frac{1}{(U^2+\bar{U}^2)^2},
\end{align} 
where $c_2=b^2/12$. 
Here we employ the stabilization condition given by 
Eq.~(\ref{eq:u2}) as discussed in Sec.~\ref{subsec:2}. 
Then the eigenvalues $(K_{\rm eig})_I$, 
and the matrix $U_{I\bar{J}}$ diagonalizing the 
above K\"ahler metric $K_{I\bar{J}}$ for 
$I,J=U^4,U^2,\tilde{T},U^3,S$ are numerically estimated in 
the case of $c_2=1$,  
\begin{align}
&(K_{\rm eig})_{U^4}\simeq \frac{4.3}{(U^2+\bar{U}^2)^2},\,\,\,
(K_{\rm eig})_{U^2}\simeq \frac{1.1}{(U^2+\bar{U}^2)^2},\,\,\,
(K_{\rm eig})_{\tilde{T}}=\frac{0.27}{(U^2+\bar{U}^2)^2},
\nonumber\\
&(K_{\rm eig})_{U^3}=K_{U^3\bar{U}^3},\,\,\,
(K_{\rm eig})_{S}=K_{S\bar{S}},
\nonumber\\
&U_{I\bar{J}}=
\begin{pmatrix}
-0.67 & -2.19 & 1 & 0 & 0\\
1 & 0.14 & 1 &0 &0 \\
-1.1 & 0.8 & 1 & 0 & 0 \\
0 & 0 &0 &1& 0 \\
0 & 0 & 0 & 0 & 1 
\end{pmatrix}
.
\end{align}

Next, we show the mass-squared matrix obtained 
from the scalar potential which is consisted of 
the K\"ahler potential~(\ref{eq:KW2}), (\ref{eq:KT}) 
and the superpotential~(\ref{eq:KW2}), (\ref{eq:gauginoW}), 
\begin{align}
m^2_{I\bar{J}}=
(K_{\rm eig})_{I\bar{K}}(U^{-1})_{\bar{K}L}V_{L\bar{M}}
U_{\bar{M}N}(K_{\rm eig})_{N\bar{J}},
\label{app:mass}
\end{align} 
where 
\begin{align}
(K_{\rm eig})_{I\bar{K}}=\delta_{I\bar{K}}/\sqrt{(K_{\rm eig})_{I}},
\end{align} 
and 
\begin{align}
V_{L\bar{M}}\simeq 
\begin{pmatrix}
V_{U^4\bar{U}^4} & 0 &0 &0 &0 \\
0 & V_{U^2\bar{U}^2} & V_{U^2\bar{\tilde{T}}} 
&0 &0 \\
0 &V_{\tilde{T}\bar{U}^2} &V_{\tilde{T}\bar{\tilde{U}}} 
& 0 &V_{\tilde{T}\bar{S}} \\
0 &0  &0 &V_{U^3\bar{U}^3} 
& V_{U^3\bar{S}} \\
0 &0 &V_{S\bar{\tilde{T}}} & V_{S\bar{U}^3} 
&V_{S\bar{S}} \\
\end{pmatrix}
.
\end{align} 
Note that here the mass-squared matrix is evaluated in the 
canonically normalized field basis ($\Phi^2,\Phi^4,\Phi^3,\Phi^S,\Phi^T$) 
with 
\begin{align}
&\bar{\Phi}^2 =\sqrt{2(K_{\rm eig})_{U^2}}
(U^{-1})_{U^2\bar{U}^{\bar{J}}}\bar{U}^{\bar{J}},\,\,\,
\bar{\Phi}^4 =\sqrt{2(K_{\rm eig})_{U^4}}
(U^{-1})_{U^4\bar{U}^{\bar{J}}}\bar{U}^{\bar{J}},\,\,\,
\bar{\Phi}^T =\sqrt{2(K_{\rm eig})_{\tilde{T}}}
(U^{-1})_{\tilde{T}\bar{U}^{\bar{J}}}\bar{\tilde{T}}
\nonumber\\
&\bar{\Phi}^3 =\sqrt{2(K_{\rm eig})_{U^3}}\bar{U}^{\bar{3}},\,\,\,
\bar{\Phi}^S =\sqrt{2(K_{\rm eig})_{S}}\bar{S},
\end{align} 
and the mass-squared matrix $m^2_{IJ}$, 
$m^2_{\bar{I}\bar{J}}$ can be obtained in the same way.

\end{document}